\documentstyle[prc,aps,epsfig,preprint]{revtex}
\tightenlines
\newcommand{\vsig}{\mbox{\boldmath $\sigma$ \unboldmath}}
\newcommand{\veps}{\mbox{\boldmath $\epsilon$ \unboldmath}}
\newcommand{\valf}{\mbox{\boldmath $\alpha$ \unboldmath}}
\newcommand{\vtau}{\mbox{\boldmath $\tau$ \unboldmath}}  
\newcommand{\vpi}{\mbox{\boldmath $\pi$ \unboldmath}}  

\begin{document}

\title{\bf Non-diffractive mechanisms in the $\phi$ meson
photoproduction on nucleons}

\author{Qiang Zhao$^1$\thanks{Email address: 
qiang.zhao@surrey.ac.uk.}, B. Saghai$^2$, and J.S. Al-Khalili$^1$}
\address{1) Department of Physics, University of Surrey, GU2 7XH, Guildford,
United Kingdom } 
\address{2) Service de Physique Nucl\'eaire, DSM/DAPNIA, 
CEA/Saclay, \\ F-91191 Gif-sur-Yvette, France}
%
%\address{3) Institut de Physique Nucl\'eaire, F-91406 Orsay Cedex, France}
%

\date{\today}

\maketitle  
  
\begin{abstract}
We examine the non-diffractive 
mechanisms in the $\phi$ meson photoproduction 
from threshold up to a few GeV using an effective Lagrangian
in a constituent quark model.
The new data from CLAS at large angles can be consistently 
accounted for in terms of 
{\it s}- and {\it u}-channel processes. 
Isotopic effects arising from the 
reactions $\gamma p\to \phi p$ and $\gamma n\to \phi n$,
are investigated by comparing  
the cross sections and polarized beam asymmetries.
Our result highlights an experimental means of 
studying non-diffractive mechanisms
in $\phi$ meson photoproduction.

\end{abstract}
\vskip 0.3cm



\vskip 0.5cm

For a long time, the study of $\phi$ meson photoproduction
has been concentrated at high energies where the diffractive 
process is the dominant source, and a Pomeron exchange 
model based on the Regge phenomenology explains the elastic $\phi$
production at small momentum transfers~\cite{donnachie}. 
In contrast with the high energy reactions, data for the photoproduction 
of the $\phi$ meson near threshold are still very sparse, 
and were available only for small momentum 
transfers~\cite{besch74}. 
The new data 
from the CLAS collaboration at JLAB~\cite{clas-phi-00}
cover for the first time momentum transfers above 1.5 (GeV/c)$^2$
with $2.66\le W\le 2.86$ GeV, 
and provide important information about mechanisms
leading to non-diffractive processes at large angles.

Initiated by the possible existence of strangeness in nucleons,
Henley {\it et al.}~\cite{henley92} 
showed that 10-20\% of strange quark 
admixture in the nucleon would result
in an $s\overline{s}$ knockout cross section 
compatible with the diffractive one near threshold.
More recently, it has been shown 
by Titov {\it et al.}~\cite{titov97-npa,titov97-prl,titov98,oh99} 
using a relativistic harmonic oscillator quark model 
that an even smaller fraction of $s\overline{s}$ of
about 5\% would produce detectable effects 
in some polarization observables. 
In Ref.~\cite{williams98}, Williams studied 
the effect of an OZI evading $\phi NN$ interaction 
by including the Born term with an effective $\phi NN$ coupling.
Quite different conclusions were drawn
from the above approaches, since 
the descriptions of the diffractive process were 
significantly model-dependent, and would 
influence not only the fraction of 
a possible $s\overline{s}$ component in the nucleon, 
but also the OZI evading $\phi NN$ coupling.
As shown in Ref.~\cite{williams98}, the $|g_{\phi NN}|$ 
could have a range of 0.3-0.8, depending on the model for  
the diffractive process.
Therefore, a reliable description of the diffractive contribution, 
which determines
what scope remains for other non-diffractive mechanisms,
is vital.
Near threshold, 
another question arising from non-diffractive 
$\phi$ meson production is
what the dominant process in the large angle
$\phi$ production might be?
In Ref.~\cite{zhao-phi-99}, we showed that OZI 
suppressed {\it s}- and {\it u}-channel contributions should be 
a dominant source for large angle $\phi$ production in 
$\gamma p\to \phi p$.

Concerning the two points noted above, 
we study here, within a quark model, the non-diffractive 
$\phi$ meson photoproduction in two isotopic channels, 
$\gamma p \to \phi p$ and $\gamma n \to \phi n$,
from threshold to a few GeV of c.m. energy.
A Pomeron exchange model, 
which was determined at higher energies,
was then extrapolated to the low energy limit with the same 
parameter. In this way, we believe 
the diffractive contribution has been reliably evaluated 
and should be a prerequisite for study of non-diffractive 
mechanisms in both reactions.
Pion exchange was also included but
found to be small.
Moreover, its forward peaking character suggests
that some other non-diffractive process is 
necessary at large angles.
An effective $\phi$-$qq$ interaction was proposed
for the {\it s}- and {\it u}-channel $\phi$ meson production, 
which will account for the large angle non-diffractive contributions
up to $W\approx 3$ GeV. 
In the quark model framework, the nucleon pole terms (Born term), 
as well as a complete set of resonance contributions 
can be consistently included. 
Our attention will be focused on the large 
angle {\it s}- and {\it u}-channel processes in this work.  
We do not take into account the strangeness 
component, although the effective $\phi$-$qq$ coupling 
might have included effects from an OZI evading process.
A comparison with 
the new data from the CLAS Collaboration
should highlight the roles played by the 
{\it s}- and {\it u}-channel $\phi$ production,
and the isotopic study will provide insight into 
any non-diffractive mechanism.

The question of whether a non-diffractive process can play
a role at a few GeV of c.m. energy, 
is still an open one. 
As pointed out
by Donnachie and Landshoff~\cite{donnachie92},
contributions from two-gluon exchanges
should be small at a few GeV, and a Pomeron exchange
would be enough.
Laget~\cite{laget00} showed that a two-gluon exchange mechanism 
might start to play a role at large $|t|$ 
with $W\approx 3$ GeV. A relatively large contribution 
was found from correlation processes. 
However, it was also shown that two-gluon exchange 
could not account for the increase in the cross sections
at large angles. A {\it u}-channel process, 
which violated the {\it s}-channel helicity conservation (SCHC), 
was then employed to explain the large angle behavior.
Interestingly,
newly submitted 
results from the CLAS Collaboration
for the $\phi$ electroproduction 
at $0.7\le Q^2\le 2.2$ (GeV/c)$^2$ and $2.0\le W\le 2.6$ GeV 
suggest that some non-diffractive mechanism
plays a role at large $t$~\cite{clas-phi-electro}.
Such results cannot generally be explained by the SCHC Pomeron 
exchange and the soft two-gluon-exchange model, but strongly imply
that some non-perturbative process might still compete
against the progressively more important perturbative QCD 
processes at a few GeV. 
To disentangle these mechanisms near threshold, 
one should start with 
those SCHC violated processes,
in particular, the {\it s}- and {\it u}-channel $\phi$ productions.
Their energy evolution to a few GeV as well as 
a measurable effect arising from 
their isotopic reaction should be 
seriously considered.

Our model consists of three processes: 
(i) {\it s}- and {\it u}-channel $\phi$ production
with an effective Lagrangian; 
(ii) {\it t}-channel Pomeron exchange; 
(iii) {\it t}-channel pion exchange.

At quark level, 
the $\phi$-$qq$ coupling is described by the effective 
Lagrangian~\cite{plb98,prc98}:
\begin{equation} \label{lagrangian}  
L_{eff}= \overline{\psi}(a\gamma_\mu +  
\frac{ib\sigma_{\mu\nu}q^\nu}{2m_q}) \phi^\mu_m \psi,
\end{equation}  
where the quark field $\psi$ can be $u$, $d$, or $s$ for the light-quark
baryon system, while $\phi^\mu_m $ represents the vector $\phi$ 
meson field.   
The 3-quark baryon system is described by the nonrelativistic 
constituent quark model (NRCQM) in the 
$SU(6)\otimes O(3)$ symmetry limit.
The vector meson is treated as an elementary point-like particle 
which couples to the constituent quark through the effective interaction.
Two parameters, $a$ and $b$, are introduced 
for the vector and tensor coupling of the $\phi$-$qq$ in the {\it s}- and 
{\it u}-channels.

At tree level, the transition amplitude from the effective Lagrangian 
can be expressed as the contributions from the {\it s}-, {\it u}- and 
{\it t}-channel processes:
\begin{equation}  
M_{fi}=M^s_{fi}+M^u_{fi}+M^t_{fi} \ .  
\label{3.1}  
\end{equation}  

In $\gamma N\to \phi N$, 
$M^t_{fi}$ vanishes since it is proportional to the charge of the
final state $\phi$ meson. 
Introducing intermediate states, 
the {\it s}- and {\it u}-channel amplitudes can be written as:
\begin{eqnarray}  
M^{s+u}_{fi}&=&i\omega_\gamma\sum_{j}\langle N_f|H_m|N_j\rangle\langle   
N_j|\frac{1}{E_i+\omega_\gamma-E_j}h_e|N_i\rangle\nonumber\\  
&&+i\omega_\gamma\sum_{j} \langle N_f|h_e\frac{1}{E_i-\omega_\phi-E_j}  
|N_j\rangle\langle N_j|H_m|N_i\rangle, 
\label{3.2}  
\end{eqnarray} 
with 
$H_m=-\overline{\psi}(a\gamma_\mu +\frac{ib\sigma_{\mu\nu}  
q^\nu}{2m_q}) \phi^\mu_m \psi$ for the quark-meson coupling vertex, and
\begin{eqnarray}  
h_e=\sum_{l}e_l{{\bf r}_l\cdot{\veps \hskip -0.16 cm }_\gamma}(1-\valf\cdot  
{\hat{\bf k}})e^{i{\bf k\cdot r}_l},~
{\hat{\bf k}}=\frac{\bf k}{\omega_\gamma},
\end{eqnarray} 
where ${\bf k}$ and $\omega_\gamma$ are the three-momentum and energy 
of the incident photon, respectively.
$|N_j\rangle$ represents the complete set of intermediate 
states. In the NRCQM, those low-lying states ($n\leq 2$)
have been successfully related to the resonances and can be 
taken into account explicitly in the formula. Higher 
excited states can be treated as degenerate in the main quantum 
number $n$ of the harmonic oscillator basis. A detailed description 
of this approach can be found in Refs.~\cite{plb98} and~\cite{prc98}. 
It should be noted that 
resonances belonging to quark model representation 
$[{\bf 70, \ ^4 8}]$ do not contribute in $\gamma p \to \phi p$ 
due to the Moorhouse selection rule at the electromagnetic 
interaction vertex~\cite{moorhouse}. 
Therefore, eight low-lying resonances will explicitly appear
in $\gamma p\to \phi p$, while there are 16 in $\gamma n \to \phi n$.

The {\it t}-channel diffractive process
 is accounted for by the 
Pomeron exchange model of 
Donnachie and Landshoff~\cite{donnachie,laget95,lee96-97}. 
In this model, the Pomeron  
mediates the long range interaction between 
two confined quarks, and behaves rather like a $C=+1$ isoscalar photon. 
We summarize the vertices as follows:

(i) Pomeron-nucleon coupling: 
\begin{equation}
F_{\mu}(t)= 3\beta_0\gamma_{\mu}f(t), \  
f(t)= \frac{(4M^2_N-2.8t)}{(4M^2_N-t)(1-t/0.7)^2} \ ,
\end{equation}
where $\beta_0$ is the coupling of the Pomeron to one light constituent quark; 
$f(t)$ is the isoscalar nucleon electromagnetic form factor 
with four-momentum transfer $t$;
the factor 3 comes from the ``quark-counting rule".

(ii) Quark-$\phi$-meson coupling: 
\begin{equation}
V_\nu(p-\frac 12 q, p+\frac 12 q)=f_\phi M_\phi\gamma_\nu \ ,
\end{equation}
where $f_\phi=164.76$ MeV is the decay constant 
of the $\phi$ meson in $\phi\to e^+ e^-$, which is
determined by 
$\Gamma_{\phi\to e^+e^-}=8\pi \alpha^2_e e^2_Q f^2_\phi/3M_\phi=1.32$ 
keV~\cite{PDG00}. 

A form factor $\mu^2_0/(\mu^2_0+p^2)$ is adopted 
for the Pomeron-off-shell-quark vertex,
where $\mu_0=1.2$ GeV is the cut-off energy, 
and $p$ is the four-momentum of the quark.
The Pomeron trajectory is
$\alpha(t)=1+\epsilon+\alpha^\prime t $, 
with $\epsilon=0.08$ and $\alpha^\prime=0.25$ GeV$^{-2}$.

The $\pi^0$ exchange is introduced via
the Lagrangian for the $\pi NN$ coupling 
and $\phi\pi\gamma$ coupling as 
\begin{eqnarray}\label{pi-nn}
L_{\pi NN}=-i g_{\pi NN}\overline\psi \gamma_5(\vtau\cdot\vpi)\psi \ ,
\end{eqnarray}
and 
\begin{eqnarray}\label{phi-pi-gamma}
L_{\phi \pi^0 \gamma}=e_N\frac{ g_{\phi\pi\gamma} }{M_\phi}
\epsilon_{\alpha\beta\gamma\delta}\partial^\alpha A^\beta
\partial^\gamma\phi^\delta\pi^0 \ .
\end{eqnarray}
Then the amplitude for the $\pi^0$ exchange can be derived in the NRCQM.
The commonly used couplings, 
${g^2_{\pi NN}}/{4\pi}= 14,
~g^2_{\phi\pi\gamma}=0.143  $, are adopted.
A sign exists between the two pion exchange amplitudes
for $\gamma p\to \phi p$ and $\gamma n\to \phi n$, 
i.e. $g_{\pi pp}=-g_{\pi nn}$, due to the isospin symmetry.

In the pion exchange, the
only parameter $\alpha_\pi=300$ MeV comes from the quark model form 
factor $e^{-({\bf q}-{\bf k})^2/6\alpha_\pi^2}$ given
by the spatial integral over the nucleon 
wavefunctions. The $\eta$ meson exchange has not been included
due to its even smaller contribution compared to the pion
exchange. A recent study~\cite{zhao-eta}
showed that the $g_{\eta NN}$ coupling 
could be as small as 1.1, which means that $\eta$ 
exchange can be neglected safely in $\phi$ meson production.

A criticism of
the application of a NRCQM to $W\approx 3$ GeV
is that relativistic effects become 
important due to the high momentum transfer between the incoming 
photon and the constituent quarks. In principle, 
one needs a relativistic version of the quark model to 
take into account the time axis. However,  
a self-consistent relativistic quark model is not available yet. 
On the other hand, the NRCQM has made impressive 
success in hadron spectroscopy as well as 
most photo-excitation helicity amplitudes for baryons~\cite{CR-00}. 
In our approach, uncertainties 
arising from NRCQM's shortcoming can be regarded
as being efficiently taken into account in two ways:
(i) The masses as well as total decay widths
of those low-lying resonances come from the experimental 
output. Therefore, one need not fit the baryon 
spectroscopy. 
(ii) A Lorentz boost factor 
for each momentum in the spatial integrals 
is employed to take into account the Lorentz contraction effects
up to $W\approx$ 3 GeV. 
In fact, it shows that energy evolution of those 
{\it s}- and {\it u}-channel terms is very important 
in relating a Pomeron exchange model to the effective 
Lagrangian model.

In the range of the CLAS measurements,
the value $|t|=2$ (GeV/c)$^2$ corresponds to a scattering 
angle of $\theta\approx 90^\circ$ in the c.m. system. 
For larger values of $|t|$, 
the cross section will reflect features from a non-diffractive
mechanism, which in our model is described by the {\it s}- and 
{\it u}-channel $\phi$ meson production.
The energy evolution as well as 
the large angle cross sections provide a direct constraint
on the parameters in our model. 
A numerical fit of the 
old data~\cite{besch74} at $E_\gamma=2.0$ GeV
and the new ones~\cite{clas-phi-00} at 3.6 GeV
gives $a=0.241\pm 0.105$ and 
$b^\prime=-0.458\pm 0.091$, which are consistent 
with previous work~\cite{zhao-phi-99}. 
Qualitatively, the ratio of parameter $a$ for the $\phi$ and 
$\omega$ meson (see Ref.~\cite{zhao-omega}) 
can be related 
to the ratio $g_{\phi NN}/g_{\omega NN}$, namely
$g_{\phi NN}/g_{\omega NN}=a(\phi)/a(\omega)$. 
In Ref.~\cite{zhao-omega}, $a(\omega)=-2.5$ 
accounted for the differential and total cross sections reasonably. 
In Ref.~\cite{nstar2001-zhao}, the best value $a(\omega)=-2.72$ 
was derived. It shows that $a(\phi)/a(\omega)=-0.096\sim -0.087$ covers 
a range very close to the value 
determined by SU(3) symmetry, 
i.e. $g_{\phi NN}/g_{\omega NN}=-\tan 3.7^\circ=-0.065$, 
where the angle $3.7^\circ$ is the deviation from the 
ideal $\omega$-$\phi$ mixing~\cite{PDG00}. 
This feature is strongly related to the effective quark-vector-meson
coupling and quark model 
phenomenology which perhaps need to be
seriously considered in future investigation.
In this work, we just treat the couplings as parameters
and leave them determined by the data. 
The signs of the parameters reflect the relative
phases between the Pomeron exchange terms and 
the {\it s}- and {\it u}-channel transition amplitudes. 
We assume that the quark-photon vertices and quark-$\phi$-meson
vertices in both the Pomeron exchange and {\it s}- and {\it u}-channel
processes have the same signs, 
even though the quark flavors are different.
Then we leave the relative phases determined by the signs of 
the parameters.
The sign for pion exchange is fixed by Eqs.~\ref{pi-nn} and ~\ref{phi-pi-gamma}.

In Fig.~\ref{fig:(1)}, the differential cross section is 
calculated at $E_\gamma=3.6$ GeV for $\gamma p\to \phi p$. 
The dot-dashed and dotted curves denote the results for 
exclusive pion exchange and pion plus Pomeron exchange, respectively.
Clearly, the Pomeron exchange is the dominant mechanism
at small momentum transfers.
It can be seen that above $|t|=2$ (GeV/c)$^2$, 
the Pomeron plus pion exchange cannot reproduce the flattened 
feature of the cross section. 
With the {\it s}- and {\it u}-channel contributions 
taken into account, the full model calculation is
presented by the solid curve. 
It is also found that 
the {\it u}-channel has a relatively stronger contribution
to the cross sections above the resonance energy region. 
Meanwhile, the {\it u}-channel nucleon pole term 
is dominant over other {\it u}-channel contributions. 
This feature is in agreement with the findings of Ref.~\cite{laget00}.
The dotted curve denotes the result excluding the
{\it u}-channel from contributing. It should be noted 
that the {\it s}- and {\it u}-channel contributions
might be slightly over-estimated since the small 
two-gluon-exchange contributions are overlooked here.

Next, we show that an isotopic 
$\phi$ meson photoproduction on the neutron will 
be able to provide us with information about the large angle 
$\phi$ meson production mechanism.

The $\phi$-$qq$ coupling in $\gamma n \to \phi n$
can be described in the same way as in $\gamma p \to \phi p$. 
But the isospin degrees of freedom distinguish 
between proton and neutron via different 
$g$-factors defined for the meson-baryon couplings~\cite{prc98}. 
Significant changes occur
due to the disappearance of the electro-interaction
in the nucleon pole terms. 
The anomalous magnetic moment of the neutron will 
result in phase change effects in the 
{\it s}- and {\it u}-channel amplitudes. 
The nucleon pole terms in $\gamma n \to \phi n$ can 
be written as
\begin{eqnarray}
M_n^s(T)&=&g_A\mu_n\frac{b^\prime}{2m_q}
\frac{M_n}{P_i\cdot k}e^{-({\bf q}^2+{\bf k}^2)/6\alpha^2}\nonumber\\
&&\times\langle \chi_f |
[ (\veps_\phi\times {\bf q})\cdot(\veps_\gamma\times{\bf k}) \nonumber\\
&&+i\vsig\cdot 
(\veps_\phi\times {\bf q})\times 
(\veps_\gamma\times{\bf k})] |\chi_i \rangle \ ,
\end{eqnarray}
for the transverse $\phi$ production in the {\it s}-channel, and 
\begin{eqnarray}
\label{Born_s_l}
M_n^s(L)&=&-ia g^t_v \mu_n\frac{M_\phi }{|{\bf q}|}
\frac{(W+M_n)}{2P_i\cdot k} e^{-({\bf q}^2+{\bf k}^2)/6\alpha^2}\nonumber\\
&&\times\langle \chi_f |\vsig\cdot 
(\veps_\gamma\times{\bf k})|\chi_i \rangle \ ,
\end{eqnarray}
for the longitudinal $\phi$ production. 
The corresponding {\it u}-channel amplitudes are
\begin{eqnarray}
M_n^u(T)&=&g_A\mu_n\frac{b^\prime}{2m_q}
\frac{M_n}{P_f\cdot k}e^{-({\bf q}^2+{\bf k}^2)/6\alpha^2} \nonumber\\
&&\times\langle \chi_f |
[ (\veps_\phi\times {\bf q})\cdot(\veps_\gamma\times{\bf k}) \nonumber\\
&&-i\vsig\cdot 
(\veps_\phi\times {\bf q})\times 
(\veps_\gamma\times{\bf k})] |\chi_i \rangle \ ,
\end{eqnarray}
for the transverse $\phi$ production, and 
\begin{eqnarray}
\label{Born_u_l}
M_n^u(L)&=&ia g^t_v \mu_n\frac{M_\phi }{|{\bf q}|}
\frac{(W+M_n)}{2W}\frac{M_n}{P_f\cdot k}
e^{-({\bf q}^2+{\bf k}^2)/6\alpha^2} \nonumber\\
&&\times\langle \chi_f |\vsig\cdot 
(\veps_\gamma\times{\bf k})|\chi_i \rangle \ ,
\end{eqnarray}
for the longitudinal $\phi$ production.
In the above equations, $\mu_n=-1/3m_q$ is the neutron's 
magnetic moment, and $m_q=330$ MeV is the constituent 
quark mass; $M_n$ and $M_\phi$ are the neutron and
$\phi$ meson, respectively; ${\bf k}$ and ${\bf q}$ are 
the momenta of the incoming photon and outgoing meson, 
respectively, while $\veps_\gamma$ and $\veps_\phi$ 
are the polarization vectors 
of the photon and meson. Two parameters, 
$a$ and $b\equiv b^\prime+a$, denote the vector 
and tensor coupling of the $\phi$-$qq$ interaction, 
and are determined in $\gamma p \to \phi p$.

An interesting feature 
related to the gauge invariance condition and arising from 
the longitudinal $\phi$ production terms
is that the separate calculation of the {\it s}- and {\it u}-channel
nucleon pole terms will result in divergence at threshold
$|{\bf q}|\to 0$. 
To get rid of such a problem, we need to add the 
{\it s}- and {\it u}-channel terms together.
Notice that $|{\bf k}|=\omega_\gamma$ in the real photon
reaction, we obtain $P_i\cdot k=|{\bf k}|W$. 
Thus, 
$M_n^{s+u}(L)$ can be written as
\begin{eqnarray}
M_n^{s+u}(L)&=&i a g^t_v \mu_n\frac{M_\phi}{|{\bf q}|}
\frac{(W+M_n)}{2W}e^{-({\bf q}^2+{\bf k}^2)/6\alpha^2} \nonumber\\
&&\times\left[-\frac{W}{\omega_\gamma(E_i+\omega_\gamma)}
+\frac{M_n}{\omega_\gamma(E_f+|{\bf q}|\cos{\theta})}\right ] \nonumber\\
&&\times \langle \chi_f |\vsig\cdot (\veps_\gamma \times {\bf k})
|\chi_i \rangle \nonumber\\
&=&-i a g^t_v \mu_n\frac{M_\phi}{|{\bf q}|}
\frac{(W+M_n)}{2W}
e^{-({\bf q}^2+{\bf k}^2)/6\alpha^2}\nonumber\\
&&\times\frac{1}{P_f\cdot k}\left[\frac{|{\bf q}|^2}{E_f+M_n}
+|{\bf q}|\cos{\theta}\right ] \nonumber\\
&&\times \langle \chi_f |\vsig\cdot (\veps_\gamma \times {\bf k})
|\chi_i \rangle \ ,
\end{eqnarray}
where $g^t_v=3$ is derived in the quark model.
In the last equation, the factor $|{\bf q}|$ in the 
denominator will be cancelled by a corresponding one
in the square-bracket, and the divergence at threshold
($|{\bf q}|\to 0$) is avoided. 
Meanwhile, the {\it u}-channel propagator
$(P_f\cdot k)^{-1}$ partly explains why the {\it u}-channel
plays an important role in the isoscalar
vector meson ($\omega$, $\phi$) photoproduction.

Using the parameters derived in $\gamma p \to \phi p$, 
the cross sections for both $\gamma p \to \phi p$ and $\gamma n \to \phi n$ are 
calculated at $E_\gamma=2.0$ GeV (Fig.~\ref{fig:(2)}). 
An obvious feature
is that the large angle cross sections
are significantly smaller for $\gamma n \to \phi n$
than for $\gamma p \to \phi p$. 
Meanwhile, a relatively stronger backward peaking is 
found from the {\it u}-channel nucleon pole term.
We also present the results without the {\it u}-channel 
contributions in Fig.~\ref{fig:(2)} (see the dotted curves). 
Comparing the dashed curves (Pomeron plus pion exchange)
to the dotted ones, we find that the {\it u}-channel 
contributions play a dominant role in both reactions. 
In another word, the {\it s}-channel resonance contributions 
are significantly smaller than the {\it u}-channel contributions
in the $\phi$ meson photoproduction. This feature, which 
has not been seen in the $\omega$ meson photoproduction, 
might make it difficult to filter signals for individual 
{\it s}-channel resonances in the $\phi$ photoproductions.
This result might be regarded as a negative result in the 
context of searching for ``missing resonances" in various 
reaction channels, however it has a positive side in that 
the forward angle kinematics might be an ideal region 
for studying the strangeness component in nucleons.
The dot-dashed curves in Fig.~\ref{fig:(2)} denote  
results for the {\it s}- and {\it u}-channel processes.

In Fig.~\ref{fig:(3)}, 
the isotopic effects 
of these two reactions 
are shown for the polarized beam asymmetry
$\check{\Sigma}$ at $E_\gamma=2.0$ GeV.
Here, $\check{\Sigma}$ is defined as 
\begin{equation}
\check{\Sigma}=\frac{\sigma_\parallel-\sigma_\perp}
{\sigma_\parallel+\sigma_\perp} \ , 
\end{equation}
where $\sigma_\parallel$ and $\sigma_\perp$
denote the cross sections for $\phi\to K^+K^-$ 
when the decay plane is parallel or perpendicular to 
the photon polarization vector.
The dashed curves represent results for 
the Pomeron plus pion exchange, which 
deviate
from $+1$ due to the presence of 
the {\it unnatural} parity pion exchange.
With the {\it s}- and {\it u}-channel contributions,
the full model calculations are denoted by the solid curves.
Explicitly, the large angle
asymmetry is strongly influenced by the presence 
of the {\it s}- and {\it u}-channel processes, while 
the forward angles are not sensitive to them.
Interferences between 
the Pomeron exchange {\it s}- and {\it u}-channel processes  
can be seen by excluding the pion exchanges (see the dot-dashed curves). 
It shows that asymmetries produced by the {\it s}- 
and {\it u}-channel processes at forward angles 
are negligible.
Since the pion exchange becomes 
very small at large angles, we conclude that
the large angle asymmetry is determined by the 
{\it s}- and {\it u}-channel processes and reflects
the isotopic effects.
The role played by the {\it s}-channel resonances
in the two reactions are presented by excluding
the {\it u}-channel contributions.
As shown by the dotted curves, the interferences from
the {\it s}-channel resonances are much weaker  
than that from the {\it u}-channel. However, they are still 
an important non-diffractive source at large angles.

It should be noted that no isotopic effects 
can be seen if only the Pomeron and pion exchange 
contribute to the cross section. 
This is because the transition amplitude 
of the Pomeron is purely imaginary, while
that of pion exchange is purely real. 
In $\check{\Sigma}$, the sign arising from the 
$g_{\pi NN}$ will disappear, which is why
the dashed curves in Fig.~\ref{fig:(3)} are the same.
We also point out that our results
for the $\check{\Sigma}$ are 
quite similar to 
findings of Ref.~\cite{titov99} at small angles, 
but very different at large angles. 
This is because only the 
nucleon pole terms for the {\it s}- and 
{\it u}-channel processes were included in Ref.~\cite{titov99}.

To show our model can be smoothly extended to a few GeV, 
we present the total cross sections 
in Fig.~\ref{fig:(4)} for both isotopic channels.
The solid and dotted curve
denote the full calculations for the 
proton and neutron reaction, respectively, 
while the dashed and dot-dashed 
curve denote the exclusive calculations of the {\it s}- and 
{\it u}-channel contributions for these two reactions, 
respectively. 
Although significant difference exists 
between the exclusive {\it s}- and {\it u}-channel
isotopic reactions, the total cross section is not 
sensitive to such an effect due to the dominance of 
Pomeron exchange.
This feature explains why such a mechanism
has in the past been neglected.

In summary, we studied the non-diffractive 
mechanisms 
in the $\phi$ meson photoproductions using 
a quark model with an effective Lagrangian in two 
isotopic channels. 
The diffractive process is accounted for by 
a Pomeron exchange model. The pion exchange 
is also included and found to be small. 
The newly published data from CLAS 
provides a good test of our model and 
highlights the mechanisms of non-diffractive 
$\phi$ production through the direct {\it s}-channel 
and crossing {\it u}-channel processes. 
The result shows that, 
up to a few GeV, these two channels
might still play a role at large angles, 
although their cross sections become small. 
Isotopic effects arising from the proton and neutron reaction
provide a means of study the {\it s}- and {\it u}-channel 
processes in experiment.
The measurement of the 
polarized beam asymmetry at large angles 
can provide detectable effects 
between these two isotopic reactions.

Conerning the search for 
signals of $s\overline{s}$ component 
in the nucleon, the forward angle kinematics 
might be selective if the findings of Refs.~\cite{titov97-prl,titov98}
are true, since at forward angles the {\it s}- and {\it u}-channel
only play a negligible role. 
Certainly, since a possible strangeness content has not been 
explicitly included in this model, the effective $\phi$-$qq$ coupling 
cannot distinguish between an OZI evading $\phi NN$ coupling
and a strangeness component in the nucleon. 
In future study, a more complex approach including 
the possible strangeness component in the nucleon will be explored.
To disentangle all the possible non-diffractive mechanisms
in $\phi$ meson photoproduction, a measurement 
of the isotopic reactions 
covering the full angle range would be also required.

%\acknowledgements
Useful comments from Z.-P. Li are gratefully acknowledged.
Q.Z. thanks M. Guidal and J.-M. Laget for valuable communications. 
We thank J.-P. Didelez, and E. Hourany 
for their interest in this work. 
The CLAS data from G. Audit are gratefully acknowledged.

%%%%%      Figure captions      %%%%%

%
\begin{figure}
%\begin{center}
%\epsfig{file=fig1.eps,height=6.cm,width=6.0cm}
%\end{center}
\caption{ Differential cross section for $\gamma p \to \phi p$
at $E_\gamma=3.6$ GeV. The dot-dashed, dashed, and solid curves
denote the pion exchange, Pomeron plus pion, and full model 
calculations, respectively, while the dotted curve 
represents full model calculation
excluding the {\it u}-channel contribution.
Data come from ~\protect\cite{besch74} (dot), 
~\protect\cite{clas-phi-00} 
(square), and ~\protect\cite{behrend78} (diamond). }
\protect\label{fig:(1)}
\end{figure}
\begin{figure}
%\begin{center}
%\epsfig{file=fig2.eps,height=5.0cm,width=9.0cm}
%\end{center}
\caption{ Differential cross section for $\gamma p \to \phi p$
and $\gamma n \to \phi n$
at $E_\gamma=2.0$ GeV. 
The dot-dashed, dashed, and solid curves
denote the {\it s}- and {\it u}-channel, Pomeron plus pion, and full model 
calculations, respectively, while the dotted curve 
represents full model calculation
excluding the {\it u}-channel contribution.
Data come from Ref.~\protect\cite{besch74}.}
\protect\label{fig:(2)}
\end{figure}
\begin{figure}
%\begin{center}
%\epsfig{file=fig3.eps,height=5.0cm, width=9.0cm}
%\end{center}
\caption{ Polarized beam symmetry
for the proton and neutron reactions at $E_\gamma=2.0$ GeV. 
The dashed, and solid curves
denote the Pomeron plus pion, and full model 
calculations, respectively, while the dotted curve 
represents full model calculation
excluding the {\it u}-channel contribution.
The dot-dashed curves denote full model calculation excluding 
the pion exchange. }
\protect\label{fig:(3)}
\end{figure}

\begin{figure}
%\begin{center}
%\epsfig{file=fig4.eps,height=6.cm,width=6.0cm}
%\end{center}
\caption{ Total cross section for $\gamma p \to \phi p$ and 
$\gamma n \to \phi n$. The solid (dotted) and dashed (dot-dashed) curves
denote the full model calculation and exclusive {\it s}- and {\it u}-channel
cross section 
for the proton (neutron) reaction, respectively.
Data come from 
Refs.~\protect\cite{crouch67,erbe68_plb,erbe68_pr,davier69,ballam73,barber82}.
See text for curve notations. }
\protect\label{fig:(4)}
\end{figure}

\begin{references}
%
\bibitem{donnachie} A. Donnachie and P.V. Landshoff,
	Phys. Lett.  B {\bf 185}, 403 (1987); 
	Nucl. Phys. {\bf B 311}, 509 (1989).
%
\bibitem{besch74} H.J. Besch {\it et al.}, 
	Nucl. Phys. {\bf B70}, 257 (1974).
%
\bibitem{clas-phi-00} E. Anciant {\it et al.}, The CLAS Collaboration,
	Phys. Rev. Lett. {\bf 85}, 4682 (2000).
%
\bibitem{henley92} E.M. Henley, G. Krein, and A.G. Williams, 
	Phys. Lett. B {\bf 281}, 178 (1992).
%
\bibitem{titov97-npa} A.I. Titov, S.N. Yang, and Y. Oh,
	Nucl. Phys. {\bf A618}, 259 (1997).
%
\bibitem{titov97-prl} A.I. Titov, Y. Oh, and S.N. Yang,
	Phys. Rev. Lett. {\bf 79}, 1643 (1997).
%
\bibitem{titov98} A.I. Titov, Y. Oh, S.N. Yang, and T. Morii,
	Phys. Rev. C {\bf 58}, 2429 (1998).
%
\bibitem{oh99} Y. Oh, A.I. Titov, S.N. Yang, and T. Morii,
	Phys. Lett. B {\bf 462}, 23 (1999).
%
\bibitem{williams98} R.A. Williams, 
	Phys. Rev. C {\bf 57}, 223 (1998).
%
\bibitem{zhao-phi-99} Q. Zhao, J.-P. Didelez, M. Guidal, and B. Saghai,
	Nucl. Phys. {\bf A660}, 323 (1999).
%
\bibitem{donnachie92} A. Donnachie and P.V. Landshoff,
	Phys. Lett.  B {\bf 296}, 227 (1992). 
%
\bibitem{laget00} J.M. Laget, 
	Phys. Lett. B {\bf 489}, 313 (2000).
%
\bibitem{clas-phi-electro} K. Lukashin {\it et al.}, The CLAS Collaboration,
	hep-ex/0101030.
%
\bibitem{plb98} Q. Zhao, Z.-P. Li and C. Bennhold, 
	Phys. Lett. B {\bf 436}, 42 (1998).
%
\bibitem{prc98} Q. Zhao, Z.-P. Li and C. Bennhold,
	Phys. Rev. C {\bf 58}, 2393 (1998).
%
\bibitem{moorhouse} R.G. Moorhouse, 
	Phys. Rev. Lett. {\bf 16}, 772 (1966).%
%
\bibitem{laget95}
	J.-M. Laget and R. Mendez-Galain,
  	Nucl. Phys. {\bf A 581}, 397 (1995).
%
\bibitem{lee96-97} 
	M.A. Pichowsky and T.-S.H. Lee, 
	Phys. Lett. B {\bf 379}, 1 (1996); 
	Phys. Rev. D {\bf 56}, 1644 (1997).
%
\bibitem{PDG00} Particle Data Group, J. Bartels, D. Haidt, and A. Zichichi,
	Eur. Phys. J. C 15, 1 (2000).
%
\bibitem{zhao-eta} Q. Zhao, B. Saghai, and Z.-P. Li, 
	nucl-th/0011069, submitted to Phys. Rev. C.
%
\bibitem{CR-00} S. Capstick and W. Roberts,
	Prog. Part. Nucl. Phys., {\bf 45} (Suppl. 2), 5241 (2000).
%
\bibitem{zhao-omega} Q. Zhao, 
	Phys. Rev. C {\bf 63}, 025203 (2001).
%
\bibitem{nstar2001-zhao} Q. Zhao, 
	Proceeding of NSTAR2001, Mar. 7-10, 2001, 
	University of Mainz, Mainz, Germany.
%
\bibitem{titov99} A.I. Titov, T.-S.H. Lee, and H. Toki,
	Phys. Rev. C {\bf 59}, R2993 (1999).
%
\bibitem{behrend78} H.-J. Behrend {\it et al.}, 
	Nucl. Phys. {\bf B144}, 22 (1978).
%
\bibitem{crouch67} H.R. Crouch {\it et al.}, 
	Phys. Rev. {\bf 156}, 1426 (1967).
%
\bibitem{erbe68_plb} R. Erbe {\it et al.}, 
	Phys. Lett. B {\bf 27}, 54 (1968).
%
\bibitem{erbe68_pr} R. Erbe {\it et al.}, 
	Phys. Rev. {\bf 175}, 1669 (1968).
%
\bibitem{davier69} M. Davier {\it et al.}, 
	Phys. Rev. D {\bf 1}, 790 (1969).
%
\bibitem{ballam73} J. Ballam {\it et al.}, 
	Phys. Rev. D {\bf 7}, 3150 (1973).
%
\bibitem{barber82} D.P. Barber {\it et al.}, 
	Z. Phys. C {\bf 12}, 1 (1982).
%
%\bibitem{Haakman} L.P.A. Hackmann, A. Kaidalov and J.H. Koch,
%  	Phys. Lett. B {\bf 365}, 411 (1996). 
%
\end{references}
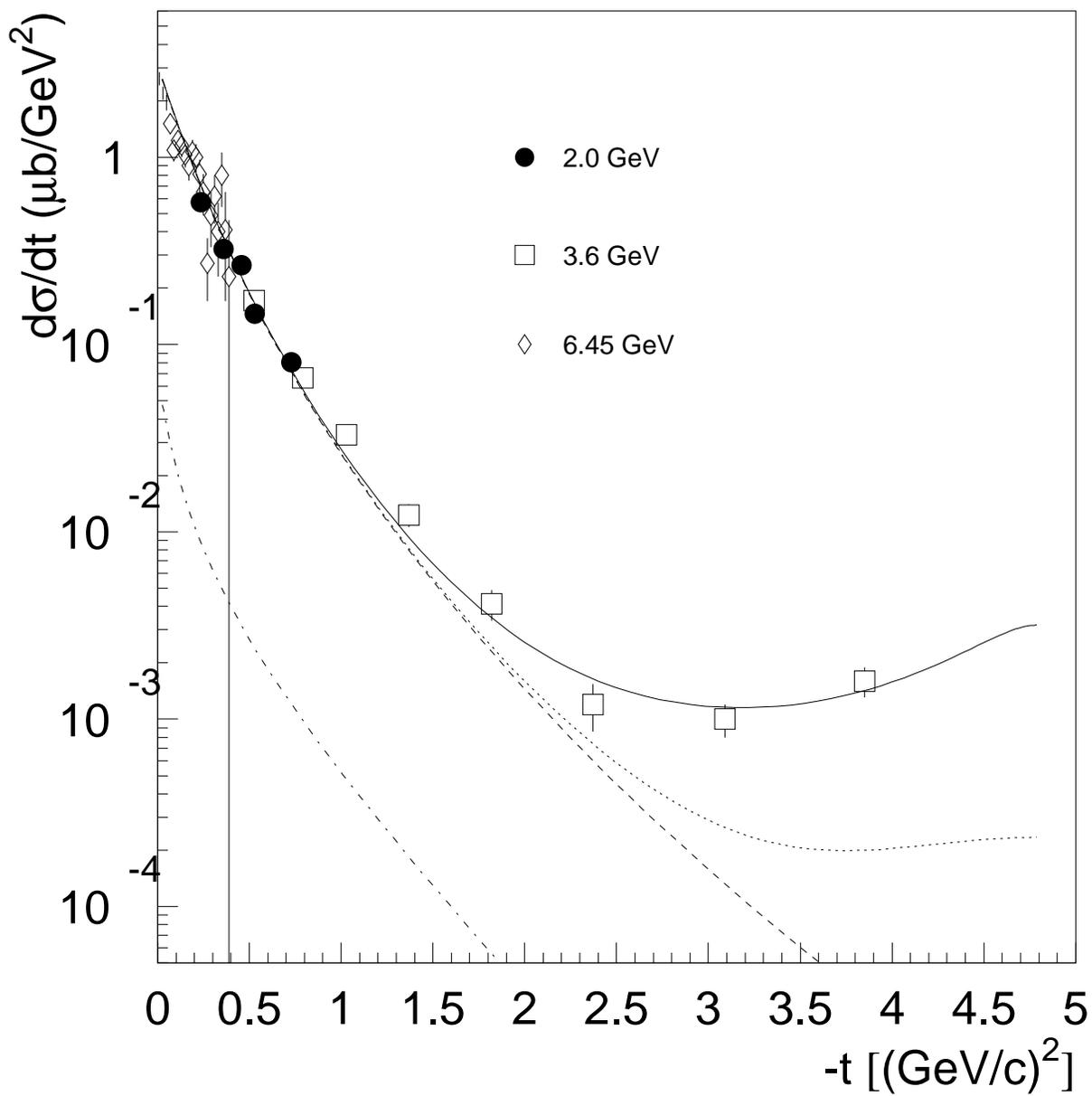
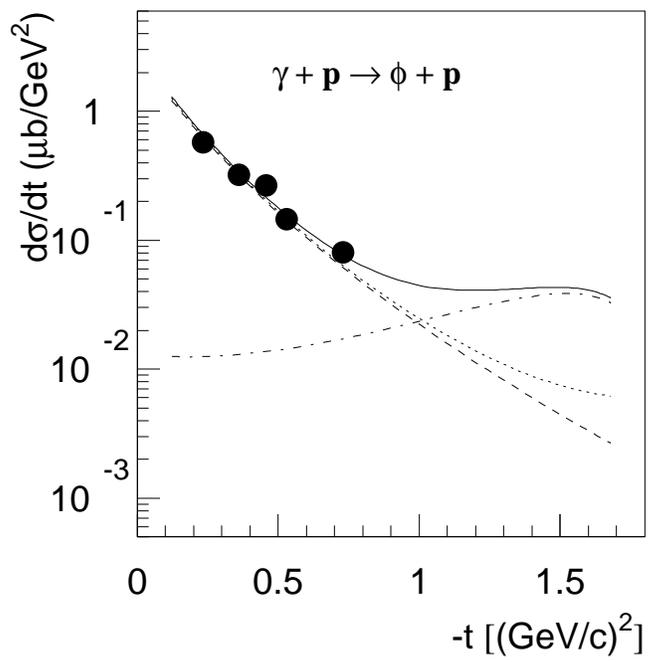
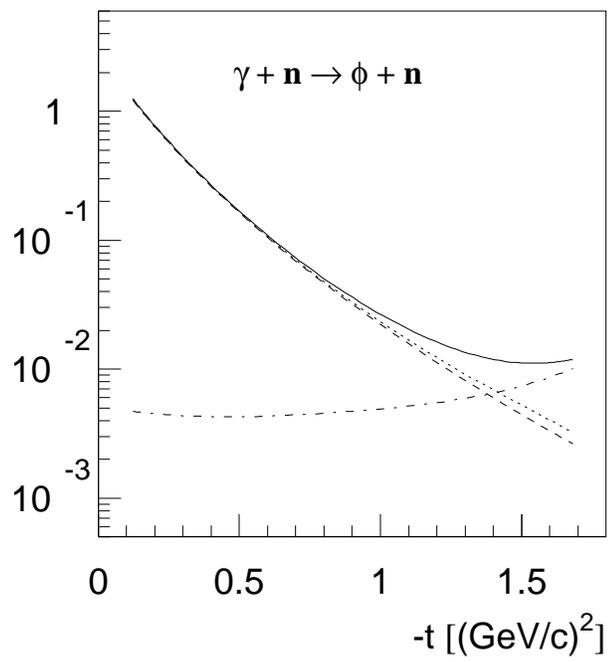
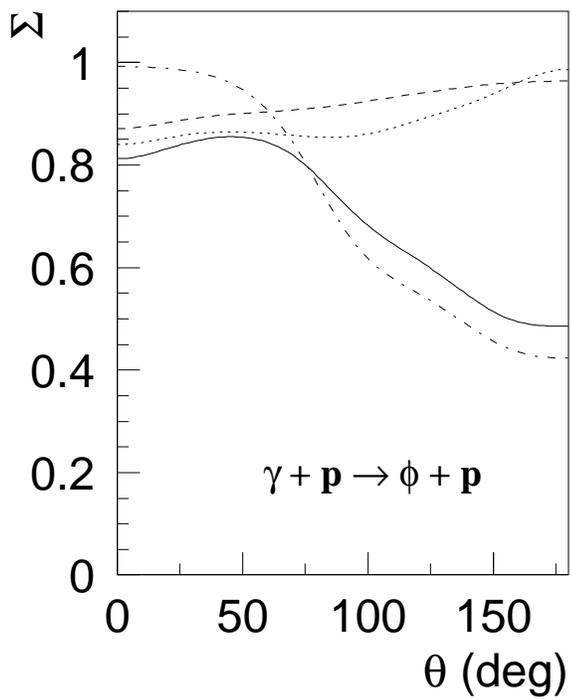
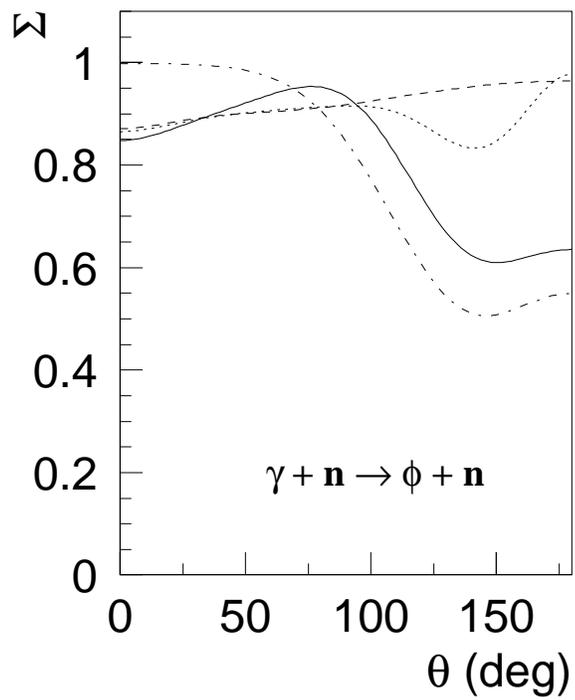
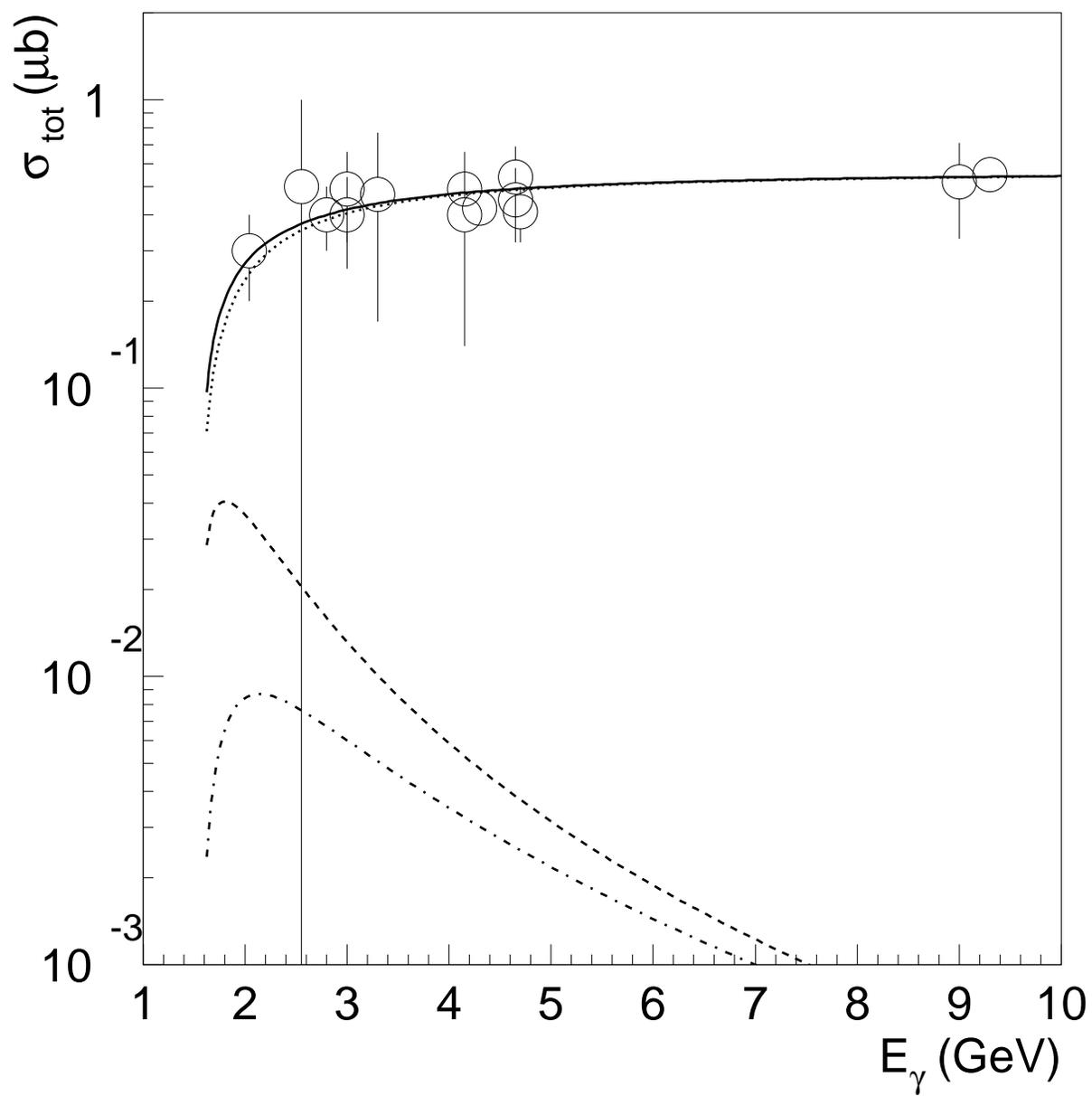
\end{document}